# Modeling of novel lateral AlGaAs/GaAs quantum well solar cell

M. Rashidi [a], Asghar Asgari [a, b*]

[a] Research Institute for Applied Physics and Astronomy, University of Tabriz, 51666-14766 Tabriz, Iran

[b] School of Electrical, Electronic and Computer Engineering, University of Western Australia, Crawley, WA 6009, Australia

**Abstract**

In this paper, a novel lateral quantum well (QW) solar cell has been introduced, and the structural parameters effects of these Nano-structures on the device's performance have been investigated. For modeling, the continuity equation has been solved in the quasi neutral regions. However, to analyze the QWs' effects, first Schrodinger – Poisson's equations have been solved self-consistently. To find the intrinsic region's absorption coefficient derived from the Fermi's golden rule, the obtained Eigen states and energies and also the effects of multilayers using Transfer Matrix Method have been employed. Then, to find the solar cell performance parameters, all radiative and non-radiative recombinations have been accounted. It is found that modifying different geometrical parameters, including the system's thickness, the wells' and barriers' width, and also structural parameters such as the barrier's mole fraction could noticeably influence the solar cell's characteristics; therefore; to obtain a good efficiency optimizing these parameters is necessary.

## 1. Introduction

In the past few years, to achieve efficiencies beyond the Shockley–Queisser limit [1] and also to reduce the cost of the existing solar cells (SCs), novel approaches have been applied to the single junction and thin film solar cells, leading to the introduction of the third generation devices. Some of these approaches[2, 3] include the use of semiconductors with different band-gaps as in multi-junction solar cells[4], introducing nanostructures as in intermediate-band cells[5]; applying selective contacts[6]; using up-conversion and down-conversion to modify the sun's polychromatic spectrum [7] ; and using the multiple carriers per single photon concept[8, 9]. In the case of quantum structured intermediate-band solar cells, quantum well (QW) solar cells[10] represents the highest efficiency nano-structured solar cell demonstrated to date[11]. In contrast to conventional homojunction cells in which sub-band-gap photons are usually lost, the use of the nanostructures would make it possible for the free carriers generated in the wells to contribute to the photocurrent alongside the high energy ones generated in the barrier regions[12]. This significantly increases the absorption capability of the structure, thus enhancing the system's Quantum efficiency[13].

The idea of using QW photovoltaic cells as a multi band gap structure was first proposed by Burnham & Duggan in 1990[14], and about one year later this idea was experimentally confirmed using a p-i-n cell through applying $GaAs/Al_xGa_{1-x}As$ multilayers as an intrinsic region[15]. From then on, the prospect of using multilayers in solar cells has become a subject of active research interest. Along these lines, numerous works, both theoretical[16-19] and experimental[20-23], have been carried out for different materials and structures. Considering the two photon absorption process available in the quantum hetrostructures, a maximum value of the quantum efficiency of about 63.2%[24] has been predicted for p-i-n type multiple QW SCs[25]. However, it has been shown that the prospect of achieving high efficiency via these two photon processes is unlikely[26] and a maximum quantum efficiency of 44.5% for a single photon absorption has been calculated[12]. In the case of experimental results, The highest single-junction QWSC power conversion efficiency achieved was 28.3% under AM1.5D [27], which was independently measured at the Fraunhofer Institute.

In these devices, to have high efficiencies, two requirements of sufficient light absorption in the wells have to be met[28], and an efficient carrier collection from the wells has to be achieved. In the typical QW solar cells, considering the fabrication issues, it is convenient to have the contacts in the growth direction. However, this would reduce the carriers' mobility, since the carriers face the barrier potential in their paths, the facility of their flow would be overshadowed and the collection of the carriers would decrease. In other words, a large number of wells required for a sufficient absorption would indeed have detrimental effects on the carriers' collection in the contacts. To overcome this problem, we have jointed the contacts laterally to the device in the investigated system, making the current direction normal to the growth direction. To model such a system, first the Schrodinger-Poisson equations were solved self-consistently to obtain the intrinsic region's absorption coefficient, which is a necessary element to find the generation in the intrinsic region. Then the continuity equation, considering both radiative and non-radiative recombination processes were solved to find the system's dark and photo currents. Finally, the effects of the system's different parameters such as the wells' and barriers' widths, and the potential well depth were investigated to find the optimized efficiency of the system.

## 2. Methods and Materials

The analyzed system is a p-i-n diode in which Quantum wells are laterally introduced to its intrinsic region. The device's length, height, and width are labeled with *l*, *h*, and *w*, respectively. The QWs have periodic structure where the wells and barriers have widths label $d_w$ and $d_b$ respectively. The incident sun-light direction, unlike usual QW solar structure, is perpendicular to the external field's direction. Other system's geometrical parameters including each regions length have been demonstrated in Fig. 1.

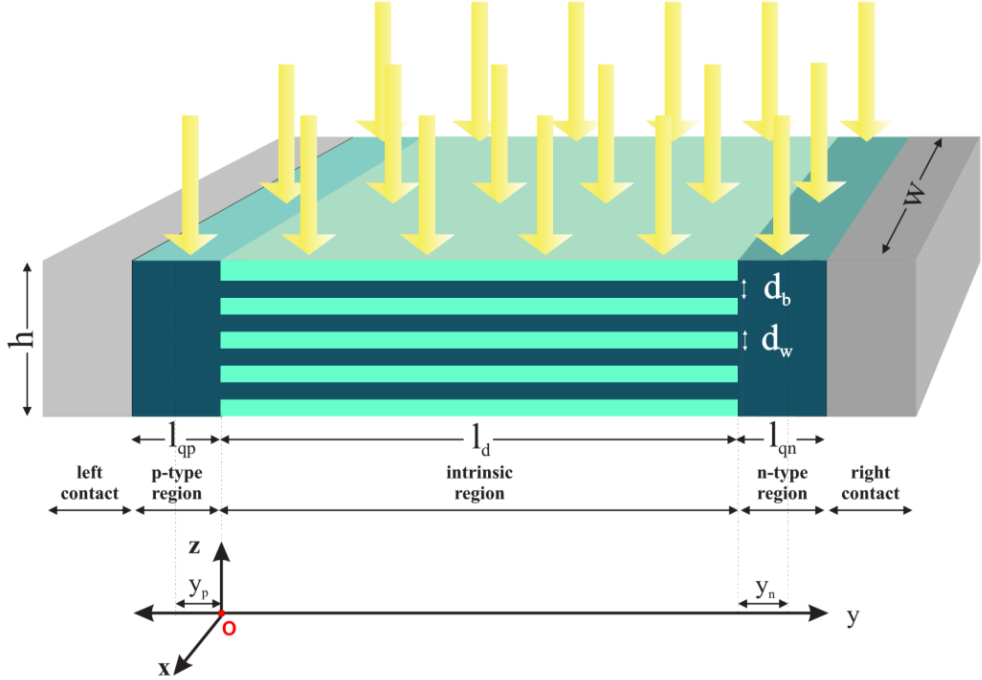

**Fig. 1 The analyzed system consisting of three regions: p-type, n-type, and the intrinsic**

To find the dark and photo currents, first, the electron and hole density profiles have been obtained through the continuity equations. It has to be noted that, in the present study, the electron and hole Fermi energies in the depletion region have been assumed to be equal to those in the adjacent quasi-neutral regions.

For the highly doped regions, considering the steady state condition and the direction of the external field, the general continuity equation simplifies to Eq. (1) for the electron in the p-type, and to Eq. (2) for the hole in the n-type regions

$$D_n \frac{d^2 n(y)}{dy^2} - U(y) + G_{l,q} = 0, \qquad (1),$$

$$D_p \frac{d^2 p(y)}{dy^2} - U(y) + G_{l,q} = 0, \qquad (2),$$

where the $U(y)$ is the net effects of thermal recombination-generation rate, the $G_{l,q}$ represents the generation rate due to light incidence in the quasi- neutral regions, $n(y)$ and $p(y)$ are the electron's and hole's densities, and $D_n$ and $D_p$ are those carriers' diffusion constants, respectively.

In the case of $G_{l,q}$ it has been presumed that the light radiation is uniform, in other words the generation is assumed to be constant on the incident plane, and the generation only changes in the *z* direction as the light penetrates into the material. Finally, the total generation of light in the *z* direction, and also in the desired frequency spectrum has been calculated and divided on each quasi- neutral region volume leading to,

$$G_{l,q} = \frac{\int_0^h \int_\lambda \varphi(\lambda) \eta_{\text{int},q} (1 - R_q(\lambda)) \alpha_q(\lambda) e^{-\alpha_q(\lambda) z} d\lambda dz}{h} \quad (3),$$

in which λ represents the input light's wavelength; $\varphi(\lambda)$ is the number of the incident photons per unit area per wavelength in AM1.5G standard solar spectra; $\eta_{\text{int},q}$, $\alpha_q(\lambda)$ and $R_q(\lambda)$ are the internal quantum efficiency, absorption coefficient constant and reflectance of the quasi- neutral regions, respectively, and finally $h$ is the device's height.

In the case of bulk quasi- neutral region's absorption coefficient ($\alpha_q(\lambda)$), Eq. (4) has been used[29],

$$\alpha_q(\lambda) = \frac{\pi q^2}{n_r c \varepsilon_0 m_0^2 \left(\frac{2\pi c}{\lambda}\right)} \rho_r\left(\hbar\left(\frac{2\pi c}{\lambda}\right) - E_g\right) \langle |\hat{e}.P_{CV}|_b^2 \rangle [f_v(k_0) - f_c(k_0)] \quad,$$

(4)

$$\rho_r\left(\hbar\left(\frac{2\pi c}{\lambda}\right) - E_g\right) = \frac{1}{2\pi^2}\left(\frac{2m_r^*}{\hbar^2}\right)^{\frac{3}{2}} \sqrt{\hbar\left(\frac{2\pi c}{\lambda}\right) - E_g} \quad, \quad k_0 = \sqrt{\frac{2m_r^*}{\hbar^2}\left(\hbar\left(\frac{2\pi c}{\lambda}\right) - E_g\right)}$$

in which $f_{v(c)}(k)$ is the Fermi-Dirac distributions for the electrons in the valence(conduction) band, $T$ represents temperature, $k_B$ and $\hbar$ are the Boltzmann and reduced Planck constants respectively, $m_r^*$ is the reduced mass which is dependent on the electron's ($m_e^*$) and hole's ($m_h^*$) effective masses, $\rho_r$ is the three-dimensional joint (reduced) density of states, $c$ is the speed of light in free space, $n_r$ is the refractive index of the material, the permittivity of the free space, $m_0$ is free electron's mass, $E_g$ is the material's band gap, $\lambda$ is the input light's wavelength, and finally $\langle |\hat{e}.P_{CV}|_b^2 \rangle$ is the average matrix element of the bulk semiconductor.

As for the $U(y)$ four main processes have been taken into consideration through the Eq. (5) to Eq. (9) [30], namely surface recombination ($U_{surf}$), band-to-band recombination ($U_{b-b}$), Shockley-Read-Hall recombination ($U_{SRH}$)[31, 32], and Auger recombination ($U_{Aug}$),

$$U(y) = U_{surf}(y) + U_{b-b}(y) + U_{SHR}(y) + U_{Aug}(y) \tag{5}$$

$$U_{Surf}(y) = \frac{p(y)n(y) - n_i^2}{\dfrac{p(y) + n_i \exp(\frac{E_i - E_{st}}{k_B T})}{S_n} + \dfrac{n(y) + n_i \exp(\frac{E_{st} - E_i}{k_B T})}{S_p}} \tag{6}$$

$$U_{b-b}(y) = b(n(y)p(y) - n_i^2) \tag{7}$$

$$U_{SHR}(y) = \frac{p(y)n(y) - n_i^2}{\dfrac{\left(p(y) + n_i \exp(\frac{E_i - E_t}{k_B T})\right)}{N_t v_{th,n} \sigma_n} + \dfrac{\left(n(y) + n_i \exp(\frac{E_t - E_i}{k_B T})\right)}{N_t v_{th,p} \sigma_p}} \tag{8}$$

$$U_{Aug}(y) = (\Gamma_n n(y) + \Gamma_p p(y))(n(y)p(y) - n_i^2) \tag{9}$$

In the above equations, $S_{n(p)}$ is the electron (hole) surface recombination velocity, $E_{st}$ is the surface trap energy, $E_t$ is the trap energy, $v_{th,n(p)}$ is the electron (hole) thermal velocity, $\sigma_{n(p)}$ is the electron (hole) capture cross section, $\Gamma_{n(p)}$ represents electron (hole) Auger coefficient, and finally $b$ is the bimolecular recombination constant.

For the boundary conditions of the used continuity equations, for the left side quasi-neutral region Eq. (10) and Eq. (11)

$$S_n [n_p - n_{p0}] = D_n \frac{dn_p}{dy} \tag{10}$$

$$n_p = n_{p0} e^{\frac{V_{bias}}{V_T}} \tag{11}$$

have been applied for the contact and depletion region interfaces, respectively, in which $V_T$ represents the thermal voltage, and in the case of the right side quasi-neutral region similar equations for the holes have been applied.

Having the carrier densities distributions, and considering no electric field effect on the minority carriers, equation Eq. (12) and Eq. (13) have been used for the minority carriers' currents in the p-region and n-region, respectively.

$$J_n(y \leq -y_p) = qD_n \frac{d(n_p - n_{p0})}{dy} \qquad (12)$$

$$J_p(y \geq y_n + l_d) = -qD_p \frac{d(p_n - p_{n0})}{dy} \qquad (13)$$

In the case of the light current, the carriers generated in the depletion and the quasi- neutral regions have been analyzed separately. Therefore, in analyzing the dark current and the quasi-neutral region's light current, it has been assumed that there is no generation or recombination in the depletion region. In other words, the current in the depletion region has been considered constant. Considering the aforementioned point, the total current through the device, without considering the depletion region's light generation, has been obtained through

$$J_{ph,q} + J_{dark} = J_n(y = -y_p) + J_p(y = l_d + y_n) \qquad (14),$$

in which $J_{ph,q}$ is the photo current density generated in the quasi- neutral region, $J_{dark}$ is the dark current density, which could be calculated independently just by replacing $G_l$ with zero. In the case of the photo-current generated in the depletion layer, the current was calculated through

$$J_{ph,d} = -q \int_{-y_p}^{l_d + y_n} G_{l,d} dy \qquad (15),$$

in which $J_{ph,d}$ is the current density generated in the depletion layer, $G_{l,d}$ is the generation in the depletion region obtained through Eq. (16).

$$G_{l,d} = \frac{\int_0^{h_{eff}(\lambda)} \int_\lambda \varphi(\lambda) \eta_{int,d} (1 - R_d(\lambda)) \alpha_d(\lambda) e^{-\alpha_d(\lambda) z} d\lambda dz}{h_{eff}} \qquad (16)$$

In Eq. (16), $\eta_{int,d}$, $\alpha_d(\lambda)$ and $R_d(\lambda)$ are the depletion region's internal quantum efficiency, absorption coefficient constant, and reflectance, respectively, and $h_{eff}(\lambda)$ is effective height that contributes to the absorption process. Unlike the quasi-natural region, the depletion layer consists of QWs making the both absorption coefficient and reflectance different. To obtain $\alpha_d(\lambda)$, the barrier regions and the regions diffused in the doping regions were treated as the quasi-neutral regions and Eq. (3) was used.

However, in the case of the well region's abruption coefficient ($\alpha_{wd}(\lambda)$), first, the Schrodinger-Poisson equations were solved self-consistently[33]. For this purpose, the Schrodinger equation was solved in the direction of the quantum well's growth, i.e. z direction, for electron (Eq. (17)) and hole (Eq. (18)), respectively.

$$\frac{-\hbar^2}{2}\frac{d}{dz}\left[\frac{1}{m_e^*(z)}\left[\frac{d}{dz}\psi_e(z)\right]\right]+E_c(z)=E_e\psi_e(z) \qquad (17)$$

$$\frac{-\hbar^2}{2}\frac{d}{dz}\left[\frac{1}{m_h^*(z)}\left[\frac{d}{dz}\psi_h(z)\right]\right]+E_v(z)=E_h\psi_h(z) \qquad (18)$$

In Eq. (17) and Eq. (18), $m_{e(h)}^*(z), \psi_{e(h)}(z)$, and $E_{e(h)}$ are the electron's (hole's) effective mass profile, wave functions, and energy; $E_{c(v)}(z)$ is the conduction(valence) band's offset.

After solving the Schrodinger equation and normalizing the resulted wave functions, Eq. (19) and Eq. (20) were used to find the electron and hole densities in the $z$ direction.

$$n(z)=\sum_i |\psi_{ei}(z)|^2 N_{is}, N_{is}=\frac{k_B T m_e^*(z)}{\pi\hbar^2}\ln\left[1+e^{(\frac{E_f-E_{ie}}{K_B T})}\right] \qquad (19)$$

$$p(z)=\sum_i |\psi_{hi}(z)|^2 P_{is}, P_{is}=\frac{k_B T m_h^*(z)}{\pi\hbar^2}\ln\left[1+e^{(\frac{E_{ih}-E_f}{K_B T})}\right] \qquad (20)$$

`In Eq. (19) and Eq. (20), $E_f$ is the Fermi energy level, $E_{ie(h)}$ and $\psi_{e(h)i}$ are the electron (hole) eigenvalues and normalized Eigen functions, respectively.

After obtaining $n(z)$ and $p(z)$, Poisson equation (Eq. (21)) were used to find $\phi(z)$ which is the electric potential created by the carriers.

$$\frac{d}{dz}(\varepsilon(z)\frac{d\phi(z)}{dz})=-\rho(z) \quad , \quad \rho(z)=|q|\left[p(z)-n(z)+N_D^+(z)-N_A^-(z)\right] \qquad (21)$$

In Eq. (21), $\varepsilon(z)$ is the permittivity profile, $\rho(z)$ is the charge distribution dependent on $N_D^+(z)$ and $N_A^-(z)$ the ionized donor and acceptor concentrations, respectively.

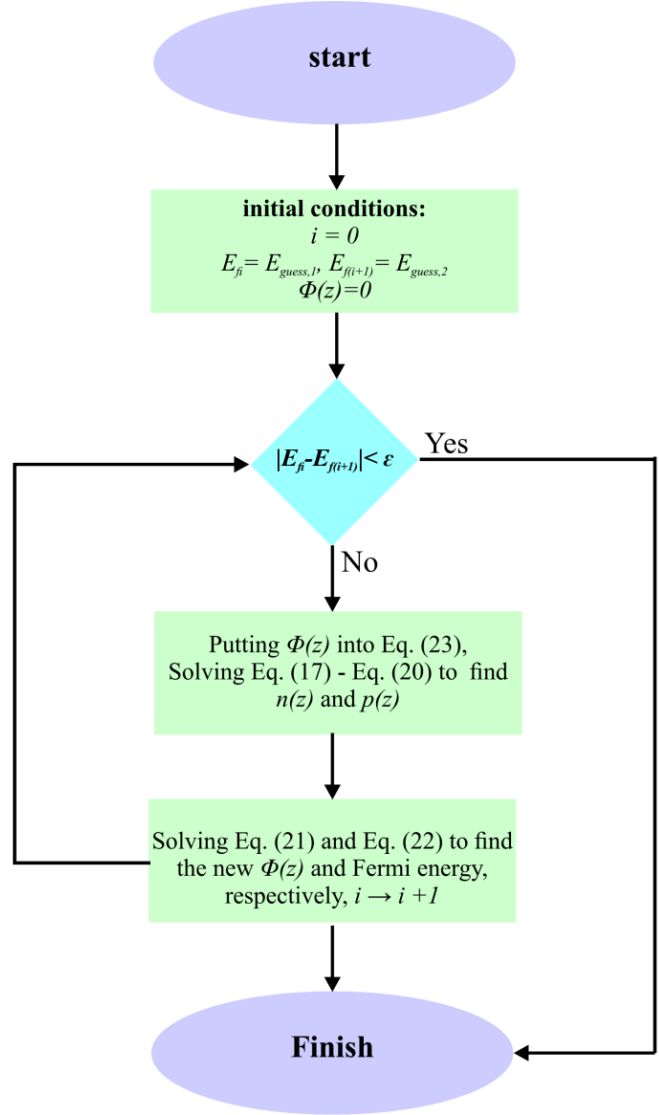

**Fig. 2.** The flow chart diagram of the scheme for solving Schrödinger-Poisson equations self-consistently

Besides finding the $\phi(z)$ through the $\rho(z)$, the new Fermi energy is calculated through the charge neutrality condition (Eq. (22))[29].

$$\int_0^h \rho(z)dz = 0 \qquad (22)$$

Finally, the obtained new Fermi energy were employed in Eq. (20) and Eq. (20) by replacing the old Fermi with the new one, and adding $\phi(z)$ to the conduction (valence) offset energy through Eq. (23).

$$E_{c(v),new} = E_{c(v),old} - |q|\phi(z) \qquad (23)$$

This cycle of Eq. (20) to Eq. (23) continues until the system's variables converge to nearly unchangeable values. The schematic of this self-consistence solution is shown in Fig. 2.

Now having the Eigen states of the quantum-wells and considering the refractive index and permittivity as $n(\omega) = n'(\omega) + in''(\omega)$ and $\varepsilon(\omega) = \varepsilon'(\omega) + i\varepsilon''(\omega)$, respectively, the absorption coefficient of the system could be calculated through [29],

$$\alpha_{wd}(\omega) = \frac{2\omega}{c} n''(\omega) \qquad (24),$$

in which c is the speed of the light in the vacuum, $n''(\omega)$ which is the imaginary part of the refractive index could be calculated through Eq. (25),

$$n''(\omega) = \frac{\varepsilon''(\omega)}{2\varepsilon_0 n'(\omega)} \qquad (25)$$

and the imaginary and real parts of the refractive index has been calculated through Eq. (26) and Eq. (27)

$$\varepsilon''(\omega) = \frac{\pi e^2}{m_0^2 \omega^2} \langle |\hat{e}.P_{CV}|^2 \rangle \left[ f_c^n(E_t = \hbar\omega - E_{hm}^{en}) - f_v^m(E_t = \hbar\omega - E_{hm}^{en}) \right] \rho_r^{2D} \sum_{n,m} |I_{hm}^{en}|^2 H(\hbar\omega - E_{hm}^{en}) \qquad (26)$$

$$n'(\omega) = \sqrt{\frac{\varepsilon'(\omega)}{\varepsilon_0}} \qquad (27)$$

In Eq. (26) and Eq. (27), $\langle |P_{CV}.\hat{e}|^2 \rangle$ is the average value of the momentum matrix element, $\rho_r^{2D}$ is two-dimensional joint (or reduced) density of states, $H$ represents the Heaviside step function, $I_{hm}^{en}$ is the overlap integral of the conduction and valence-band envelope functions, and finally $f_c^n - f_v^m$ is the Fermi-Dirac population inversion factor which in the case of un-pumped semiconductor is nearly equal to 1. In the case of average momentum matrix element, for the quantum well structures, unlike the bulk materials, is polarized dependent. The real part of the permittivity ($\varepsilon'(\omega)$) in Eq. (27) could be obtained through Kramers-Kronig relations and Eq. (26) which has been mentioned in Eq.(28).

$$\varepsilon'(\omega) = \varepsilon_0 - \frac{2e^2}{m_0^2} \langle |\hat{e}.P_{CV}|^2 \rangle \left[ f_c^n(E) - f_v^m(E) \right] \rho_r^{2D} \int_0^\infty \sum_{n,m} \frac{|I_{hm}^{en}|^2 H(\hbar\omega - E_{hm}^{en})}{\omega'(\omega^2 - \omega'^2)} d\omega' \qquad (28)$$

About $R_d(\omega)$ in Eq. (16), respectively, the Transfer Matrix Method(TMM)[34] has been applied. As a result, for system like Fig. 3 the relation between incoming wave amplitude ($A_0$) and the outgoing (transmitted) amplitude ($A_S$) could be related through,

$$\begin{pmatrix} A_0 \\ B_0 \end{pmatrix} = \begin{pmatrix} M_{11} & M_{12} \\ M_{21} & M_{22} \end{pmatrix} \begin{pmatrix} A_s \\ 0 \end{pmatrix} \qquad (29)$$

$$\begin{pmatrix} M_{11} & M_{12} \\ M_{21} & M_{22} \end{pmatrix} = D_0^{-1} \left[ \prod_{i=1}^{N} D_i P_i D_i^{-1} \right] D_s \qquad (30)$$

in which $D_\alpha$ is the dynamic matrix of each region and $P_i$ is the propagation matrix of *ith* layer.

Finally, having the relation between $A_0$ and $A_s$, the transmittances of any multilayer structure such as the depletion layer and quasi neutral regions ($N=1$) could be obtained through,

$$R = \left| \frac{M_{21}}{M_{11}} \right|^2 \qquad (31).$$

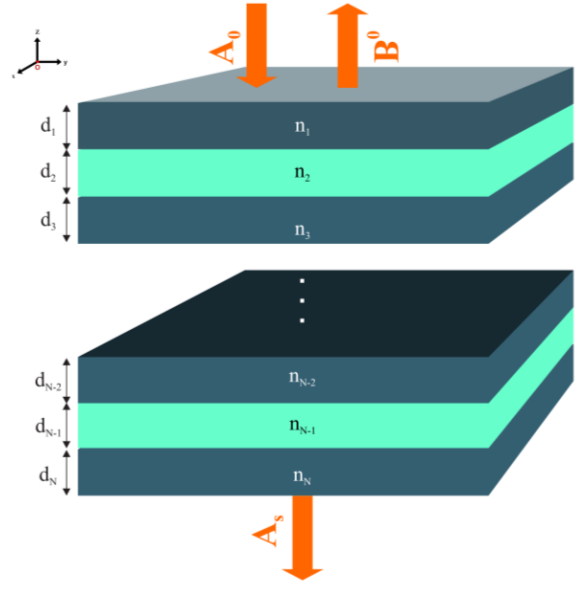

**Fig. 3. The investigated multilayer dielectric medium**

## 3. Results & Discussion

Here the model has been applied to a particular p-i-n diode made up of GaAs semiconductor in which $Al_xGa_{1-x}As$ is the barrier introduced in the intrinsic region to make the hetrostructure. Different characteristics of the device such as photocurrent, dark current, and absorption coefficient of the device have been obtained. Subsequently, the effects of device's different parameters including the number of the quantum wells (or the system's height ($h$)), the aluminum content $x$, the width of the quantum wells ($d_w$), and the width of the barrier ($d_b$) have been analyzed.

### 3.1. Structural definition

The general properties of the materials formed the device (i.e., GaAs and $Al_xGa_{1-x}As$) have been summarized in Table 1 . In the case of the doping concentration of the regions, it has been assumed that the n-region and p-region are uniformed doped with $N_{A,p} = N_{D,n} = 10^{18}(1/cm^3)$, respectively, and the intrinsic region doping has chosen nearly intrinsic, i.e. $N_{D,i} = 10^7(1/cm^3)$. The ambient temperature has been assumed 300k, and unless the effects of the device's parameters haven't modified, $x$ has been assumed 0.3.

**Table 1 Material characterizations of GaAs & Al$_x$Ga$_{1-x}$As**

|  | *GaAs* | *Al$_x$Ga$_{1-x}$As* | *Ref* |
|---|---|---|---|
| *Electron affinity (ev)* | 4.07 | 4.07-1.1x     x<0.45 <br> 3.64 - 0.14x     x>0.45 | 35, 36 |
| *Gap energy (ev)* | 1.42 | 1.42+1.247x     x<0.45 <br> 1.9 + 0.125x+0.143x$^2$     x>0.45 | 36, 37 |
| *Intrinsic carrier concentration (1/cm$^3$)* | 2.11×10$^6$ | $2.1 \times 10^5$     $x < 0.1$ <br> $2.1 \times 10^3$     $0.1 < x < 0.4$ <br> $2.5 \times 10^2$     $0.4 < x < 0.7$ <br> $4.3 \times 10^1$     $0.7 < x < 1$ | 38 |
| $m_e^* / m_0$ | 0.067 | 0.067+0.0838x     x<0.45 | 36 |
| $m_h^* / m_0$ | 0.50 | 0.62+0.14x | 37, 36 |
| $\varepsilon / \varepsilon_0$ | 13.2 | 13.18-3.12x | 37, 36 |

The transportation and recombination properties of the quasi-neutral regions, which have been made up of GaAs has been summarized in Table 2.

**Table 2 transportation and recombination properties of the quasi-neutral regions**

| parameter | value | parameter | value | Ref |
|---|---|---|---|---|
| *Γ$_n$ (cm$^6$/s)* | 1.3×10$^{-30}$ | *S$_n$ (cm/s)* | 450 | 39 |
| *Γ$_p$ (cm$^6$/s)* | 6.5×10$^{-30}$ | *S$_p$ (cm/s)* | 450 | 39 |
| *b (cm$^3$/sec)* | $(3 \times 10^{-10})(300/T)^{\frac{3}{2}}(E_{g-GaAs}(ev)/1.5)^2$ | *E$_t$ (ev)* | $3E_g/4$ | 40 |
| *N$_t$(1/cm$^3$)* | 10$^{11}$ | *E$_{st}$(ev)* | $3E_g/4$ | - |

## 3.2. Geometrical definition

The geometrical characteristics of the analyzed system to obtain the device's characteristics have been summarized in Table 3. It has to note that, in the case of analyzing system's different parameters some of these parameters (including *h, d$_w$, and d$_b$*) might have been modified.

**Table 3 Geometrical parameters of the analyzed system**

| **w** | 500 nm | **l$_{qp}$= l$_{qn}$** | 10 nm |
|---|---|---|---|
| **l** | 1000 nm | **d$_w$** | 5 nm |
| **h** | 200 nm | **d$_b$** | 5 nm |

## 3.3. Results
### 3.3.1. Obtaining the solar cell's characteristics

Solving the Schrodinger – Poisson self-constantly the Eigen states of the QWs in the *z* direction have been presented in Fig. 4. This figure shows that only one mini-band have been created both in the conductance and valance bands of the supper lattice.

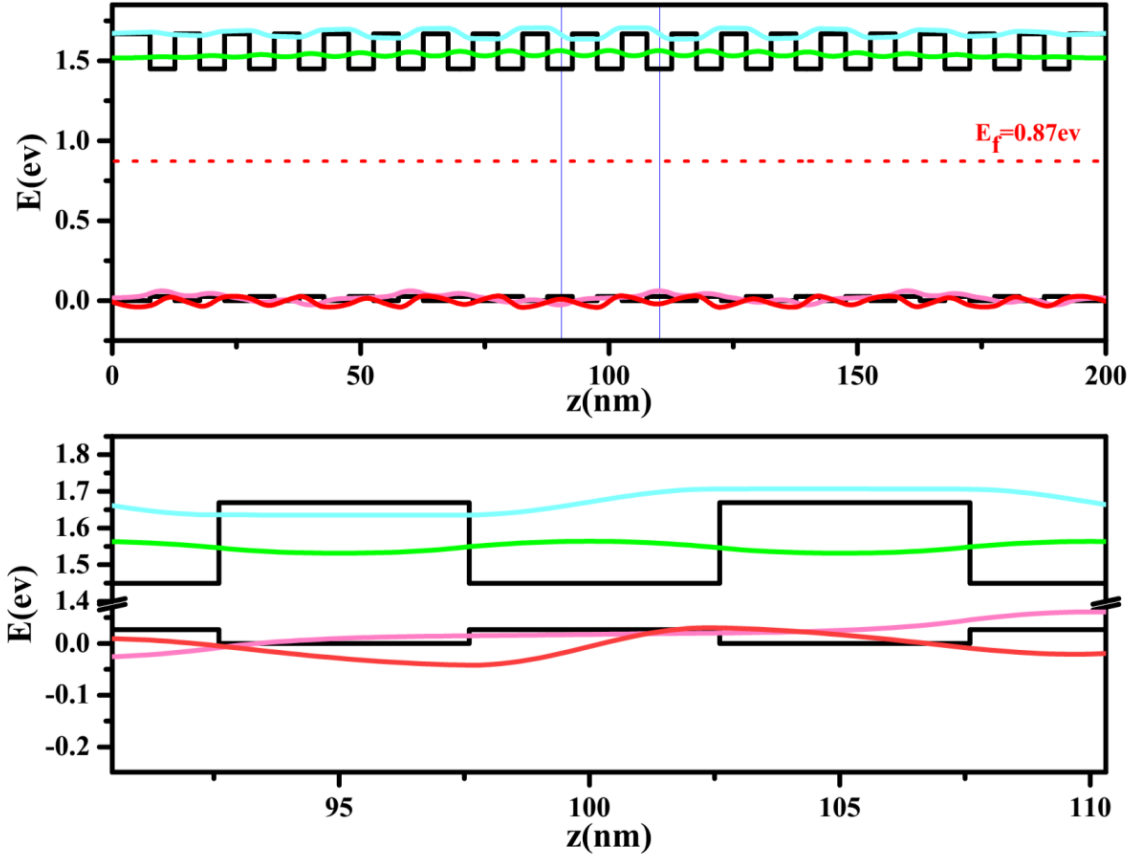

**Fig. 4 The Eigen states of the QWs in the z direction.**

The charge carriers confined in the QWs (*n(z)* and *p(z)*) have been shown in Fig. 5. As it can be seen from Fig. 5, the charge carrier is relatively low which has to do with the low doping level of the intrinsic region. The potential ($\phi(z)$) of the total charge distribution ($\rho(z)$) is relatively small (around $10^{-8}$(v)) which means that $\phi(z)$ doesn't disturb the material dependent potential of the conduction ($E_c$) and valence band ($E_v$). Furthermore, As shown in Fig. 5 (left) the free electron density (*n(z)*) is higher than hole ones (*p(z)*) which has much to do with the reason that the intrinsic region's impurities have been assumed to be donor type. (potential)

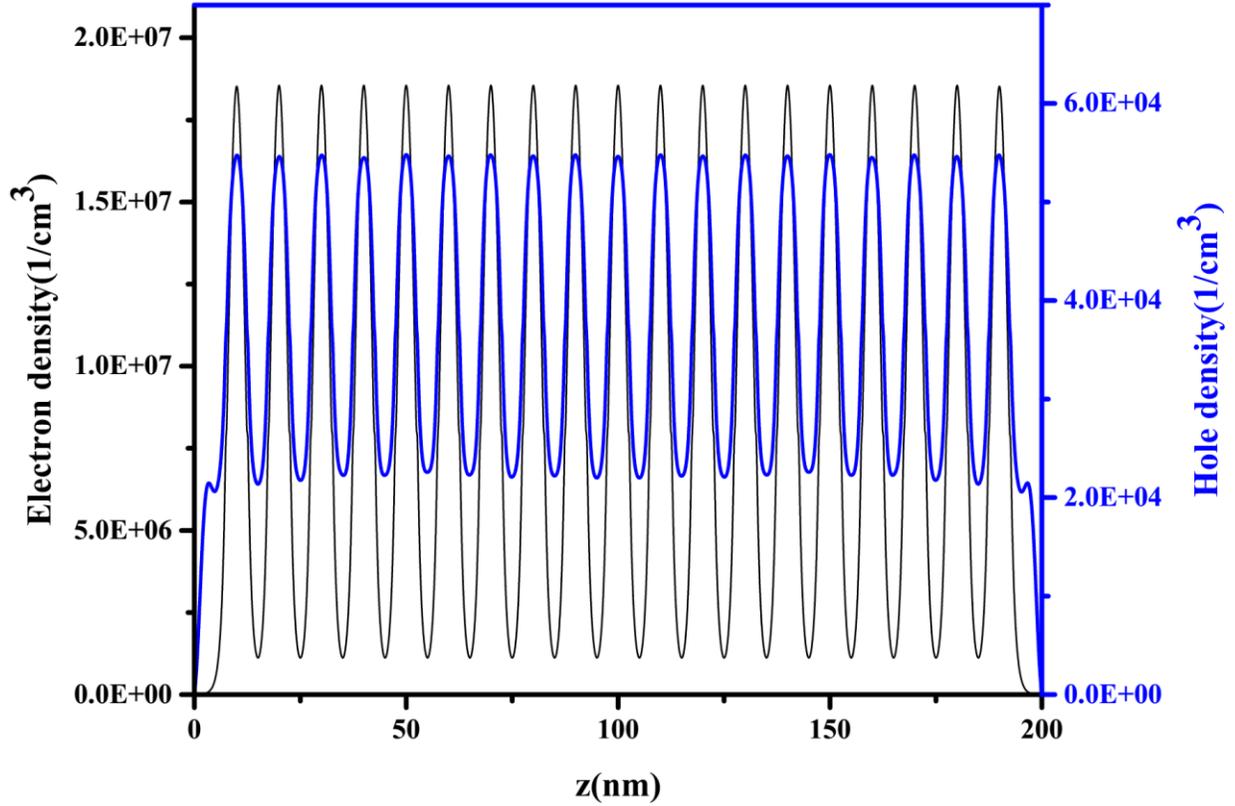

**Fig. 5 (left) the charge carriers confined in the QWs (n(z) and p(z)), (right) potential created by charge distribution**

In the next step, having the eigen states in hand, the absorption coefficient of the intrinsic region has been calculated. As shown in Fig. 6, the absorption of the bulk regions, including the barrier and quasi-neutral regions, have increased parabolically; however, the wells' absorption coefficient shows discontinuous nature arising from the inherent discontinuity of these region's density of states. Comparing the absorption coefficient of GaAs in the quasi-neutral regions and quantum wells, the results show a noticeable enhancement in the absorption coefficient of the quantum well regions resulting from the quantization of the electron's momentum in the z direction. Moreover, this comparison shows that in QWs the absorption cut-off energy is higher than quasi-neutral regions ones, which is reasonable due to the incensement of the gap energy in these quantum structures. The reason behind the differences in the n-region and p-region of the quasi-neutral regions has to do with the difference of those regions' Fermi levels. These results show that the introduction of the QWs could enhance the absorption capacity of the device.

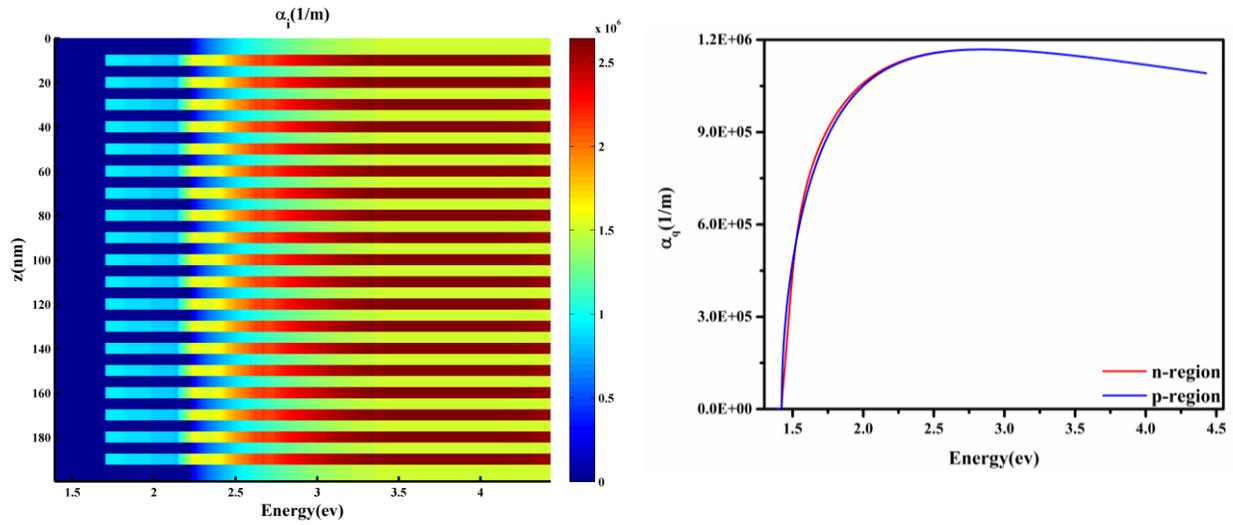

**Fig. 6 The absorption coefficient of different regions of the device: (left) intrinsic region, (right) quasi-neutral region**

Using the TMM, the reflectance of the intrinsic region have been calculated. As shown in Fig. 7, the system's reflectance for some wave-lengths is almost zero which shows that in some wave lengths due to destructive interference of the reflected lights reflecting from the multi-layered system.

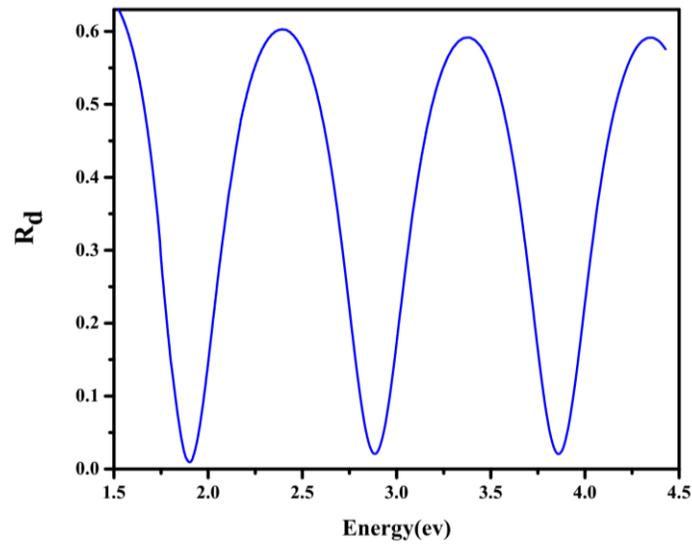

**Fig. 7 The quasi-neutral regions and intrinsic regions' reflectance**

Having the reflectance and absorption coefficient of the three regions, the photo-current of the device have been calculated. The system's dark and total current have been shown in Fig. 8.

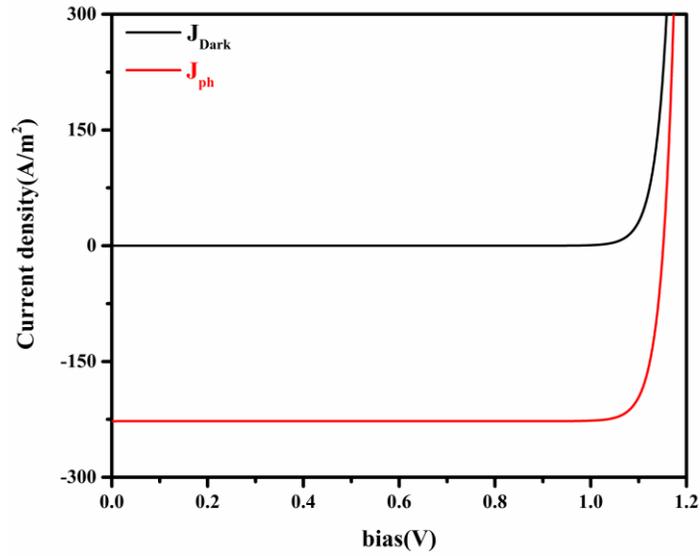

**Fig. 8 The system's dark and total current**

Finally, the solar cell's characteristics have been calculated and summarized in Table 4 .

**Table 4 The analyzed solar cell's parameters**

| $I_{sc}$ (µA) | $V_{oc}(V)$ | $\eta\%$ | FF |
|---|---|---|---|
| 22.74 | 1.1529 | 4.68 | 0.89 |

### 3.3.2. The effect of solar cell's parameters
### 3.3.2.1. QW's number ($N_w$)

Increasing the number of the QWs would increase the height ($h$) of the device, and therefore, increase the number of the absorbed photons; however, it also increases the device's volume. As a result, the optical generation and short current density reduce ($J_{sc}$). On the other hand, the device's efficiency increases which is due to short current ($I_{sc}$) increase. The effect of this parameter has been depicted in Fig. 9.

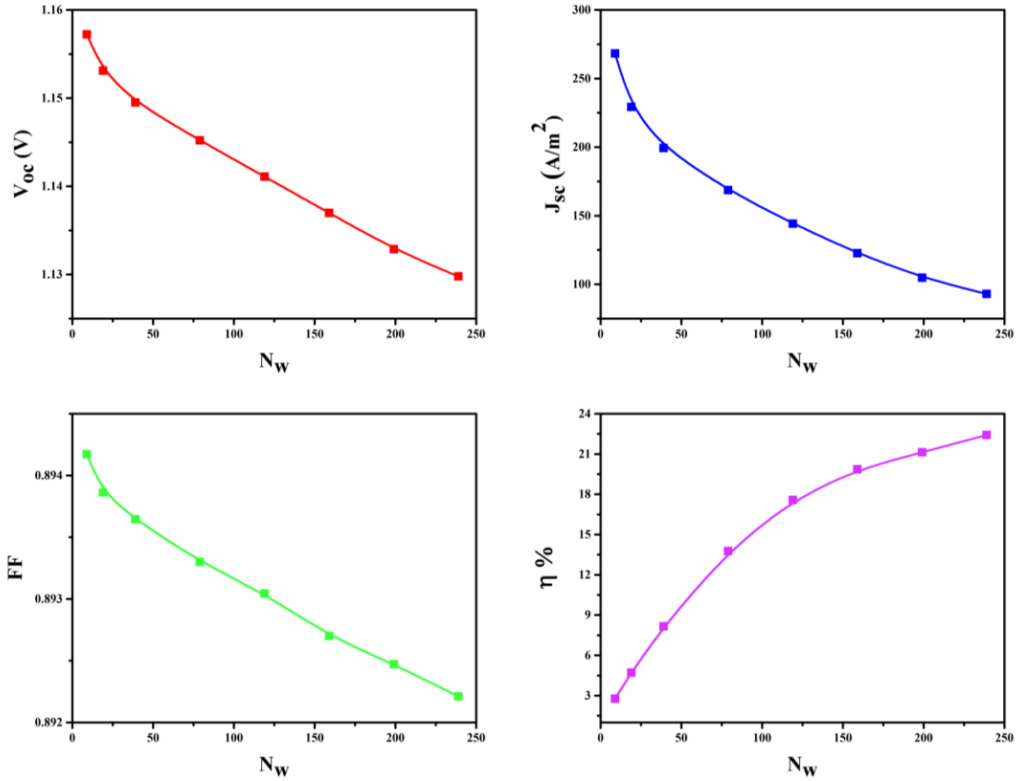

Fig. 9 The effect of the number of the wells on the device's parameters

### 3.3.3. Aluminum content ($x$)

Changing $x$ would actually modify $Al_xGa_{1-x}As$ material characteristics (Table 1), and therefore, modify its absorption coefficient. For example, increasing x would increase the material's DOS and band gap, which the first effect would increase, and the second one would decrease the efficiency of the device. Furthermore, changing the Aluminum content would modify the barrier depth both in CB and VB which would actually affect the device's performance. In other words, as we increase $x$, the barrier depth increases as well, therefore, the wells show a higher quality of confinement which increases the QWs' absorption coefficient. Fig. 10 shows the effect of the Aluminum content on two systems with different heights ($h$ =400nm, 1000nm). The figure shows that at relatively low Aluminum content the device's efficiency decreases through increasing x; however, the device's behavior goes the reverse trend in higher Aluminum content. The device with the smaller height ($h$ =400nm), below x=0.3 there is only one subband in the well, and the effects which decreases the efficiency of the device, including increase in both $Al_xGa_{1-x}As$ band gap and the subband's energy, dominates the device's performance. However, as x goes higher than the transition point (x=0.3) the number of the states in the well region increase making the quantum confinement effects more effective. In the case of the device with the higher height (*h* =1000nm), the transition point is shifted to smaller values (x=0.2). This arises from the fact that as the number of the wells increases the number of the confined states increases well, thus the system becomes more predisposed to lose its dependency on classical properties. Moreover, the effect of the Aluminum content increase for the values smaller than the transition points is kind of similar for both devices. However, for values larger than the transition points, the effect of Aluminum content variation on the device's efficiency is higher for the thicker device. This is in agreement with the differences in the dominant source of the two regions which is more sensitive to $N_w$ for the device with the higher thickness.

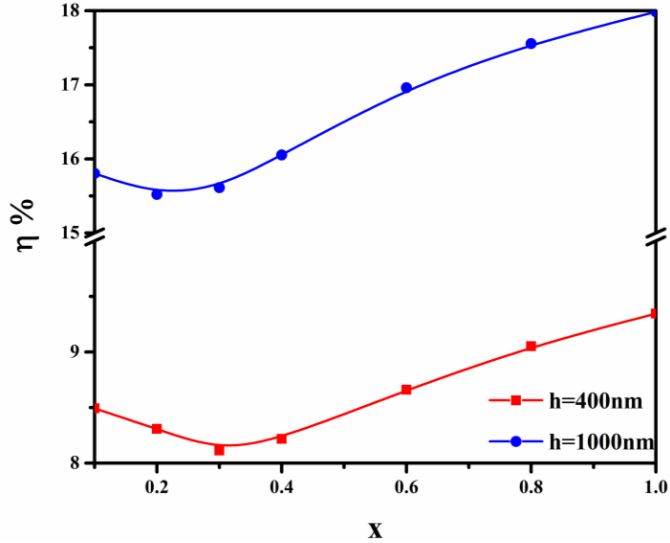

**Fig. 10 The effect of the x fraction on the device's parameters**

### 3.3.4. The QWs' width ($d_w$)

Here the effects of the QWs' width ($d_w$) on the device's performance has been analyzed and the results have been presented in Fig. 11. For this purpose, as analyzing the Al content effect the model has been applied for the two systems with different heights and all other parameters, except $d_w$ which is swept and *x* which is chosen 1, have remained as the first analyzed one. It is expected that increasing $d_w$ could affect device's performances through various aspects. First, as $d_w$ is increased, the effective height ($h_{eff}=N_w d_w$) would also increase. This would decrease the QWs' reduced DOS ($\rho_r^{2D}$) leading to the decrease of the QWs' absorption coefficient. Moreover, increasing this parameter would influence the energies of the states, especially the ones which are kind of confined in the well regions. For example, it could lower Eigen states energy leading to both increase in the number of the confined states and decrease of the well's band gap which would increase the system's device's efficiency. The other effect arising from the increase of $h_{eff}$ is the effect that this modification has on the portion that the well and barrier regions contribute to the current cross section. As Fig. 11 shows, considering all these parameters, increasing the well's width would decreases the device's efficiency.

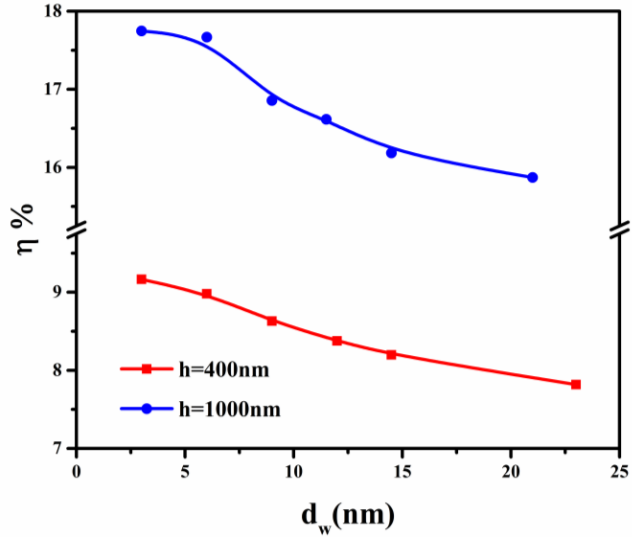

**Fig. 11 The effect of the well's width ($d_w$) on the solar cell's parameters**

### 3.3.5. The Barrier's width ($d_b$)

Here the effect of the barrier's width on the device's performance has been investigated. All the parameters of the system are as the previously analyzed one, except the $d_w$ has been assumed 5nm and $d_b$ has been swept. Increasing $d_b$ at some aspects is like decreasing $d_w$, so generally it might be believed to have the opposite trend of decreasing $d_w$. However, as

Fig. 12 shows, there are some reasons making these two seemingly same modifications different.

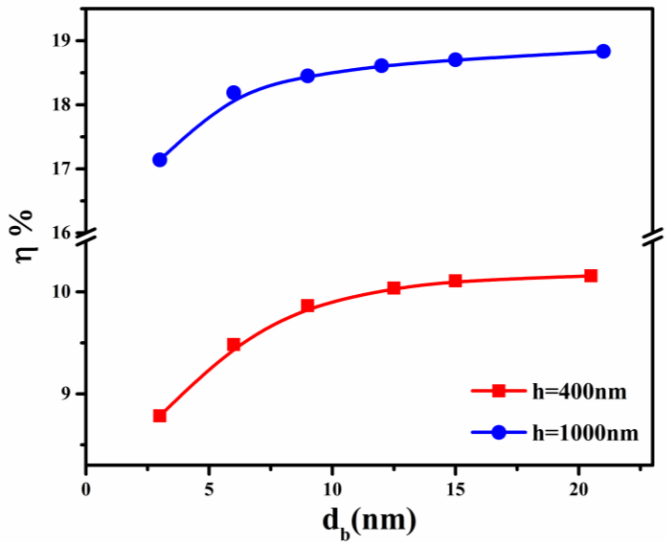

**Fig. 12 The effect of the barriers' width ($d_b$) on the solar cell's parameters**

These differences arise from the different effects that these modifications have on the QWs' eigen-states. For example, increasing $d_w$ would decrease the eigen-values of the QWs' states, or increasing $d_b$ would decrease the degree of the coupling of the adjacent QWs' Eigen-states.

# 4. Conclusion:

In this paper, a novel laterally Quantum-well solar cell has been introduced. In these systems, because of transport of carriers in the sheets (with les scattering), the solar cell's efficiency would be higher. Moreover, the results show a great dependency of the Quantum-well solar cell's efficiency on system's parameters. In the matter of geometrical parameters, increasing Quantum wells' width or decreasing the barriers' width would noticeably decrease the system's efficiency. Beside geometrical parameters, structural parameters could also change the system's efficiency. For instance, as the barrier material's mole fraction increases the device's efficiency usually increases, as well.